\documentclass[aps,pra,reprint,superscriptaddress,longbibliography
]{revtex4-1}
\usepackage{amsmath,amssymb,graphicx,cleveref}
\begin{document}

\title{Experimentally quantifying the advantages of weak-value-based metrology}

\author{Gerardo I. Viza}
\email{gerviza@pas.rochester.edu}
\author{Juli\'{a}n Mart\'{i}nez-Rinc\'{o}n}
\affiliation{Department of Physics and Astronomy and Center for Coherence and Quantum Optics, University of Rochester, Rochester, New York 14627, USA}
\author{Gabriel B. Alves}
\affiliation{Instituto de F\'{i}sica, Universidade Federal do Rio de Janeiro, 21941-972 Rio de Janeiro, Rio de Janeiro, Brazil}
\author{Andrew N. Jordan}
\affiliation{Department of Physics and Astronomy and Center for Coherence and Quantum Optics, University of Rochester, Rochester, New York 14627, USA and}
\affiliation{Institute for Quantum Studies, Chapman University, 1 University Drive, Orange, California 92866, USA}
\author{John C. Howell}
\affiliation{Department of Physics and Astronomy and Center for Coherence and Quantum Optics, University of Rochester, Rochester, New York 14627, USA}

\begin{abstract} We experimentally investigate the relative advantages of implementing weak-value-based metrology versus standard methods. While the techniques outlined herein apply more generally, we measure small optical beam deflections both using a Sagnac interferometer with a monitored dark port (the weak-value-based technique), and by focusing the entire beam to a split detector (the standard technique).
By introducing controlled external transverse detector modulations and transverse beam deflection momentum modulations, we quantify the mitigation of these sources in the weak-value-based experiment versus the standard focusing experiment.
The experiments are compared using a combination
of deterministic and stochastic methods. In all cases, the weak-values technique performs
the same or better than the standard technique by up to two orders of magnitude in precision for our parameters. We further measure the statistical efficiency of the
weak-values-based technique. By postselecting on $1\%$ of the photons, we obtain $99\%$ of the available Fisher information of the beam deflection parameter.

\end{abstract}

\maketitle

\section{Introduction}\label{Intro}
Weak-value amplification is a metrological technique intended to precisely measure small parameters, such as optical beam deflections~\cite{Dixon2009}, phase shifts~\cite{StarlingPhase,Jayaswal}, frequency shifts~\cite{StarlingFreq}, velocities~\cite{Viza}, and temperature~\cite{Egan}. A weak-value technique was shown by Viza \emph{et al.}~\cite{Viza} to saturate the Cram\'{e}r-Rao bound for small velocity measurements. Quantum mechanically, it consists of a weak interaction of a system with a meter, separated in time by nearly orthogonal pre- and post-selection measurements on a system~\cite{Aharonov}. In this technique, the parameter of interest controls the weakness of the interaction. As such, a small shift in the value of the parameter corresponds to a large shift in the meter.

A well-designed weak-values experiment concentrates almost all available information about the parameter of interest into the small fraction of events that survive the post-selection process~\cite{Hofman,Jordan2013,Decoherence,Bie}, except for a negligibly small amount that can in principle be extracted from the non-post-selected events.  Existing experiments of this kind also have a wave optics interpretation so long as we focus on intensities and not photon counts~\cite{Howell2010}.

In previous works, Hosten and Kwiat measured an angstrom beam shift to detect the spin Hall effect of light~\cite{Kwiat}. Subsequent experiments demonstrated the ability to measure down to $400$ frad of deflection~\cite{Dixon2009}, and also showed gains in the signal-to-noise Ratio (SNR)~\cite{StarlingOpti}. 
Weak-value amplification has also been shown to improve the SNR relative to the non-post-selected case in the presence of additive correlated technical noise by the Steinberg group in Ref.~\cite{Steinberg}.

The question of quantifying the relative advantages of weak value metrology techniques has taken on a renewed importance. Several recent theoretical papers claimed that weak-value amplification shows no advantages in comparison with techniques that use all the photons when optimal statistical estimators are used~\cite{Knee2013,Ferrie2013,Tanaka,Zhu,Walmsley}. When considering ideal, quantum-limited experiments and detectors
this was shown a number of years ago by the authors in Ref.~\cite{StarlingOpti}. However, in the presence of certain kinds of technical noise sources, assuming statistically independent photons, we have claimed theoretically in a recent paper~\cite{Jordan2013} that when using {\it optimal} statistical estimators that saturate the Cram\'{e}r-Rao bound, the weak-value amplification method can give advantages in the estimation of a parameter compared to other methods. That paper gave several predictions in this regard. Since these kinds of technical noise sources plague every kind of metrological experiment, a way of approaching the fundamental quantum limits in their presence is of great interest. There have been numerous papers claiming advantages for weak-value-based metrology (see, e.g.,~\cite{BrunnerSimon,KedemNoise,DresselRecycle}), including a series of very recent works~\cite{Omar,Span1,kedem,Pang,Xiadong,jayaswal2014,Jayaswal}.

\begin{figure}[t!]
 \centering
 \includegraphics[scale=0.38]{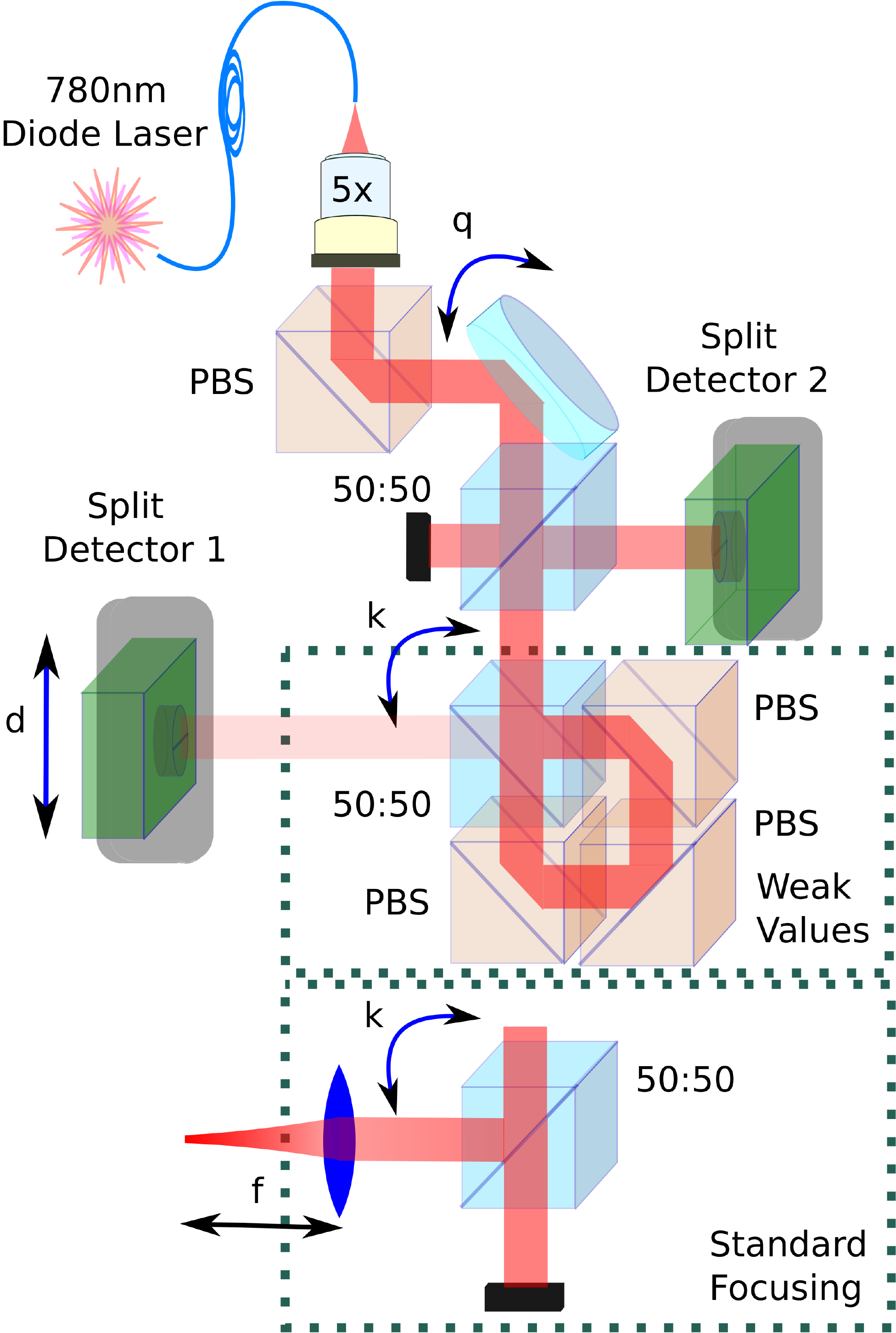}
 \caption{(Color online) (Experimental schematic) We use a c.w. Gaussian beam exiting a fiber. We compare the different experiments WVT (upper box) and the ST (lower box) to determine a beam deflection using split detector $1$ (SD$1$). The WVT uses a Sagnac interferometer, and the ST focuses the deflected beam with focal length $f$. The piezo $50$:$50$ beam splitter imparts a momentum kick $k$, which we determine from the beam shift on split detector $1$. The two external modulations are labeled as $q$ and $d$. The $d$ refers to the transverse detector modulation and $q$ refers to the transverse momentum modulation. Polarizing Beam Splitters (PBS) are in orange and $50$:$50$ beam splitter are in blue. The PBS work as mirrors given that the light is vertically polarized. The split detector $2$ (SD$2$) is used to collect the bright port beam shift from the WVT.}
 \label{fig:setup}
\end{figure}

In this paper, we present data to quantify the advantages weak-value-based experiments offer for optical beam deflection measurements. Weak values, unlike eigenvalues, can be complex. In this work we focus on imaginary-weak-value experiments, which previous studies have shown to perform better than real weak values for metrology in some circumstances~\cite{BrunnerSimon}. The imaginary weak value corresponds to a shift in one variable resulting in the displacement of its conjugate. In our experiment, this indicates that a momentum shift results in a transverse beam displacement. 
Similarly, in the standard technique, after the momentum kick a focusing lens effects a Fourier transform on the beam which results in a transverse beam displacement.
In making the comparison between the weak-value-based technique (WVT) and the standard technique (ST), we pay special attention to the statistical estimators used.  For the ST, we use an estimator which can achieve the lowest possible variance for unbiased estimators. We make a detailed study of the efficiency of the WVT statistical estimator. Figure~\ref{fig:setup} contains diagrams of the experiments carried out.
We begin with a Sagnac interferometer to measure a beam deflection as in~\cite{Dixon2009,StarlingOpti} and add two external modulating sources, meant to simulate noise sources at a given frequency: a transverse momentum modulation, $q$, and a transverse detector modulation, $d$ (see Fig.~\ref{fig:setup}).
We define a measure of sensitivity of the experiment to these modulations to be the ratio, $\mathcal{R}$, of the signal to the external modulation amplitudes.
Using single-frequency external modulations, we show that the WVT performs as well as or better than the ST and the amount of advantage is governed by the geometry and choice of parameters of the experiment.

In what follows, we show modulations and noise sources outside the interferometer of the WVT are un-amplified and thus suppressed compared to the signal, while the ST responds similarly to all modulations and noise sources. In the experiments we perform, all modulating sources are independent of the parameter of interest. This holds true even for naturally occurring laser-beam-jitter noise.
In demonstrating these effects, we report the following results. (i) The ratio $\mathcal{R}$ of the WVT indicates that the transverse momentum, $q$, and transverse detection, $d$, modulations are suppressed over ST. (ii) Comparing the deviation in measurements of transverse momentum $k$ to the smallest predicted error, the WVT offers improvement for both transverse momentum and transverse detection modulations over the ST. (iii) We show there is practically no Fisher information lost to the bright port of the WVT. (iv) Lastly, the WVT suppresses {\it naturally} occurring laser-beam-jitter noise over the ST. For all our results, we use the same acquisition time for both the WVT and the ST.

The paper is organized as follows. In Sec.~\ref{Theory}, we review the theory of the beam deflection metrology based on the WVT and the ST. We also review the concepts of Fisher information and the Cram\'{e}r-Rao bound applied to these experiments. In Sec.~\ref{Experiment}, we describe the experimental setups. In Sec~\ref{Results1}, we present a comparison of the WVT and ST based on accuracy and deviation of beam deflection measurements. In Sec.~\ref{Results2}, we show the efficiency of estimations using the Fisher information. In Sec.~\ref{Results3}, we compare WVT and ST with naturally occurring intrinsic laser beam jitter. Lastly in Sec.~\ref{Conclusions}, we give the conclusions we draw from these experiments.

\section{Theory}\label{Theory}
We consider the experimental setup shown in Fig.~\ref{fig:setup}, hereafter referred to as the WVT (upper box) or ST (lower box).
In the weak-values protocol, a
Gaussian beam of radius $\sigma$ with initial electric field transverse profile,
$E_{in}(x)=E_0\exp\left(-x^2/4\sigma^2\right)$, is sent through a
Sagnac interferometer. The beam enters the interferometer through a piezo-actuated $50$:$50$ beam splitter (BS), which imparts a momentum kick, $k$, and phase, $\phi$, to the reflected beam. The phase, $\phi$, is given by a constant deflection in the vertical $y$ axis. The beams recombine and interfere back at the BS. The recombination of the beams entangles the which-path degree of freedom to our position-momentum degree of freedom~\cite{Dixon2009}. Two output fields exit the BS as in Howell \emph{et al.}~\cite{Howell2010}:
\begin{equation}
E_{out}(x) =
\left( {\begin{array}{c}
\sin(-kx+\phi/2)\\
\cos(-kx+\phi/2)
\end{array} } \right)E_0e^\frac{-x^2}{4\sigma^2}.
\end{equation}
We assume the momentum kick $k$ is small for the weak interaction approximation, $k^2\sigma^2\cot^2(\phi/2)\ll1$. Expanding the trigonometric functions up to $\mathcal{O}(x)$, we re-exponentiate and complete the square to arrive with the dark and bright port beam shifts. Then the intensity profile takes the form:
\begin{equation}
I^{wv}_{out}(x) = I_0
\left( {\begin{array}{c}
\sin^2(\frac{\phi}{2})\exp\left[-(x+\delta_{d})^2/2\sigma^2\right]\\
\cos^2(\frac{\phi}{2})\exp\left[-(x-\delta_{b})^2/2\sigma^2\right]
\end{array} } \right),
\label{Eq:wvOut}
\end{equation}
where the dark and bright port shifts are given by $\delta_{d}=2\sigma^2k\cot(\phi/2)$ and $\delta_{b}=$ $2\sigma^2k\tan(\phi/2)$ respectively. The superscripts ${wv}$ and ${st}$ refer to the WVT and ST respectively.

In the ST protocol, we consider a lens in order to optimize this technique for deflection measurements. As shown in Fig.~\ref{fig:setup}, the lens with a focal length $f$ focuses the beam on Split Detector $1$ (SD$1$). The SD then measures the transverse displacement $fk/k_0$. The intensity profile of the ST is written as:
\begin{equation}
I^{st}_{out}(x) = I_0\exp{\left[\frac{-1}{2\sigma_{f}^2}\left(x-f\frac{k}{k_0}\right)^2\right]},
\label{Eq:stOut}
\end{equation}
where the beam radius at the focus is $\sigma_f = f/2k_0\sigma$~\cite{Jordan2013}, and  $k_0=2\pi/\lambda$ is the wave number defined by the center wavelength of the laser $\lambda$.
\begin{table}[h]\caption{A summary of the detection techniques following the theory described, where $k,d,$ and $q$ are the momentum kick of interest, the transverse detector modulation and the momentum kick from transverse momentum modulation respectively. The beam shift is given by $\delta x$, and the distance from the external modulating mirror, $q$, to the detector SD$1$ is given by $L$.
}
\begin{ruledtabular}
\begin{tabular}{ccc}
\textrm{Sources}&
\textrm{Weak-values tech.}&
\textrm{Standard tech.}\\
\colrule
$k$ & $\delta x_k=2k\sigma^2\cot(\phi/2)$ & $\delta x_k=fk/k_0$\\
$d$ & $\delta x_d=d$ & $\delta x_d=d$\\
$q$ & $\delta x_q=Lq/k_0$ & $\delta x_q=fq/k_0$\\
\end{tabular}
\label{tab:table1}
\end{ruledtabular}

\end{table}

For both techniques, we use single-frequency external modulations of two conjugate domains of our experiments: a deflecting mirror with transverse momentum modulation $q$, and the SD$1$ on a stage with a transverse detector modulation $d$. Since Gaussian white noise can be modeled by randomly changing the size of the modulation, one can add each Fourier component independently and expect similar results to Ref.~\cite{Jordan2013}.

Now we compare the WVT and the ST when measuring a momentum kick $k$ in the presence of external modulations. 
We quantify the size of the signal in comparison to the background modulation with ratio $\mathcal{R}$. The ratio $\mathcal{R}$ is the beam shift at the detector $\delta x$ due to the signal $k$, divided by the modulation $q$ or $d$ (values from Tab.~\ref{tab:table1}), $\mathcal{R}_{q,d}=\delta x_k/\delta x_{q,d}$. For the two modulations, we find,

\begin{subequations}
\begin{equation}
  \mathcal{R}^{wv}_{d}=\frac{\delta x^{wv}_k}{\delta x^{wv}_d}=\frac{2k_0 \sigma^2}{f}\cot(\phi/2)\mathcal{R}^{st}_{d},
\label{Eq:GeoD}
\end{equation}
\begin{equation}
  \mathcal{R}^{wv}_{q}=\frac{\delta x^{wv}_k}{\delta x^{wv}_q}=\frac{2k_0 \sigma^2}{L}\cot(\phi/2)\mathcal{R}^{st}_{q},
\label{Eq:GeoQ}
\end{equation}
\label{Eq:Geo}
\end{subequations}

\noindent
where the superscripts, ${wv}$ and ${st}$ refer to the technique in use---either the WVT or the ST respectively. From Eqs.~(\ref{Eq:Geo}), $\mathcal{R}^{st}\ll\mathcal{R}^{wv}$ holds true for reasonable values of $\sigma$, $L$, $f$, and $\phi$. We will show this explicitly in Sec.~\ref{Results1}. We note, the analysis here uses the dark port of the WVT in Eq.~(\ref{Eq:wvOut}).

We also compare the WVT to the ST using the Fisher information~\cite{b1,b2}.   
Knowing the transverse probability distribution in the presence of random fluctuations arriving on the SD$1$ allows us to calculate the Fisher information, $\mathcal{I}(k)$, with respect to the momentum kick $k$.
The Fisher information sets the minimum possible statistical variance using unbiased estimators, called the Cram\'{e}r-Rao bound, $\mathcal{I}^{-1}$. (For a more complete theory of Fisher information see \emph{e.g.,} Jordan \emph{et al.}~\cite{Jordan2013}.) The Fisher information can be written as

\begin{equation}
\mathcal{I}(k)=\int dx\,P(x;k)\,\left[\frac{\partial}{\partial k}\ln{P(x;k)}\right]^2,\label{Eq:FisherInformation}
\end{equation}
where $P(x;k)$ is of the normalized form of Eqs.~(\ref{Eq:wvOut}) or~(\ref{Eq:stOut}). 
$P(x;k)$ is the probability distribution of the photon arriving on the detector with transverse momentum $k$.
The Fisher information assumes discrete events---although the light intensity was derived in Eq~(\ref{Eq:wvOut}) or~(\ref{Eq:stOut}). 
With Eq.~(\ref{Eq:FisherInformation}), we arrive at the Fisher information with respect to the momentum kick $k$ and number of photons $N$ (independent trials) for our two techniques:
\begin{subequations}
\begin{equation}
  \mathcal{I}^{wv}_{Dark}(k)=4N\sigma^2\cos^2(\phi/2),\label{Eq:WvDark}
\end{equation}
\begin{equation}
  \mathcal{I}^{wv}_{Bright}(k)=4N\sigma^2\sin^2(\phi/2),\label{Eq:WvBright}
\end{equation}\label{Eq:FisherWV}
\end{subequations}
\begin{equation}
  \mathcal{I}^{st}(k)=4N\sigma^2\label{Eq:StFisher}.
\end{equation}
Where in Eq.~(\ref{Eq:FisherWV}) the Fisher information, $\mathcal{I}$, of the dark and bright ports of the WVT are denoted by the subscripts ${Dark}$ and ${Bright}$ respectively.

The two Fisher informations for the WVT arise because of the two exit ports of the BS as in Eq.~(\ref{Eq:wvOut}). Adding the Fisher information from each port leads us to the total Fisher information found in the ST~\footnote{We note that our split detectors reduce the available Fisher information by a factor of $2/\pi$ and increase the error by $\sqrt{\pi/2}$ for both the WVT and ST; nevertheless, for most of our analysis this factor dropped out, but we included it in the calculations when needed.}. We note both the ST and the WVT transform deflections into displacements in conjugate bases with the Fisher information proportional to the beam waist before the transformation. We also note that Eq.~(\ref{Eq:StFisher}) can also be found from the quantum Fisher information~\cite{Bie,Tanaka} derived from the transverse wave-function, which gives the same result.  
We also note that the Fisher information results in Eqs.~(\ref{Eq:FisherWV}) are only valid for the weak-interaction approximation, $k^2\sigma^2\cot^2(\phi/2)\ll1$.

\section{Experiment}\label{Experiment}
We use a grating feedback laser with $\lambda\approx780$ nm coupled into a polarization-maintaining single mode fiber. The Gaussian mode exits the fiber, reflects through a polarizing beam splitter (PBS) for polarization purity, and reflects off of a piezo-actuated mirror $q$ (see Fig.~\ref{fig:setup}).

In the WVT, the beam propagates through the piezo-mounted $50$:$50$ BS and enters a Sagnac interferometer of three PBS acting as mirrors for the vertically polarized light. The beam recombines back in the piezo-mounted $50$:$50$ BS and exits through the dark and bright ports. The photons exiting the dark port are sent to SD$1$ on a piezo-actuated translation stage. To collect the bright port photons, we add an extra $50$:$50$ BS before the interferometer to direct them to SD$2$ as in Fig.~\ref{fig:setup}.

For the ST, the Gaussian beam is reflected from the $50$:$50$ BS. Then, the beam is focused onto the SD$1$ on a piezoactuated translation stage as in Fig.~\ref{fig:setup}.

We calibrated the piezoresponses independently by reflecting the beam from the actuated devices to the SD$1$. The piezoresponses of the actuated $50$:$50$ BS, the piezo-actuated mirror, and the piezo-actuated translation stage were calibrated to be $\alpha_1\approx 68.6$ pm/mV, $\alpha_2\approx 31.6$ pm/mV, and $\alpha_3\approx 75.8$ pm/mV, respectively. The piezo calibrations differ because of different materials and loads.


\section{Results---Comparison of the two Techniques}\label{Results1}
First, we will show that modulating sources external to a weak values amplifying system are not amplified and can thus be suppressed. It is important to note in the WVT the modulations external to the interferometer arrive at the detector without amplification and with a reduced number of photons. While the ST uses a lens to focus the beam with every modulation (external and the source) to the detector with all the photons.

\begin{figure}[t!]
  \centering
  \includegraphics[scale=0.16]{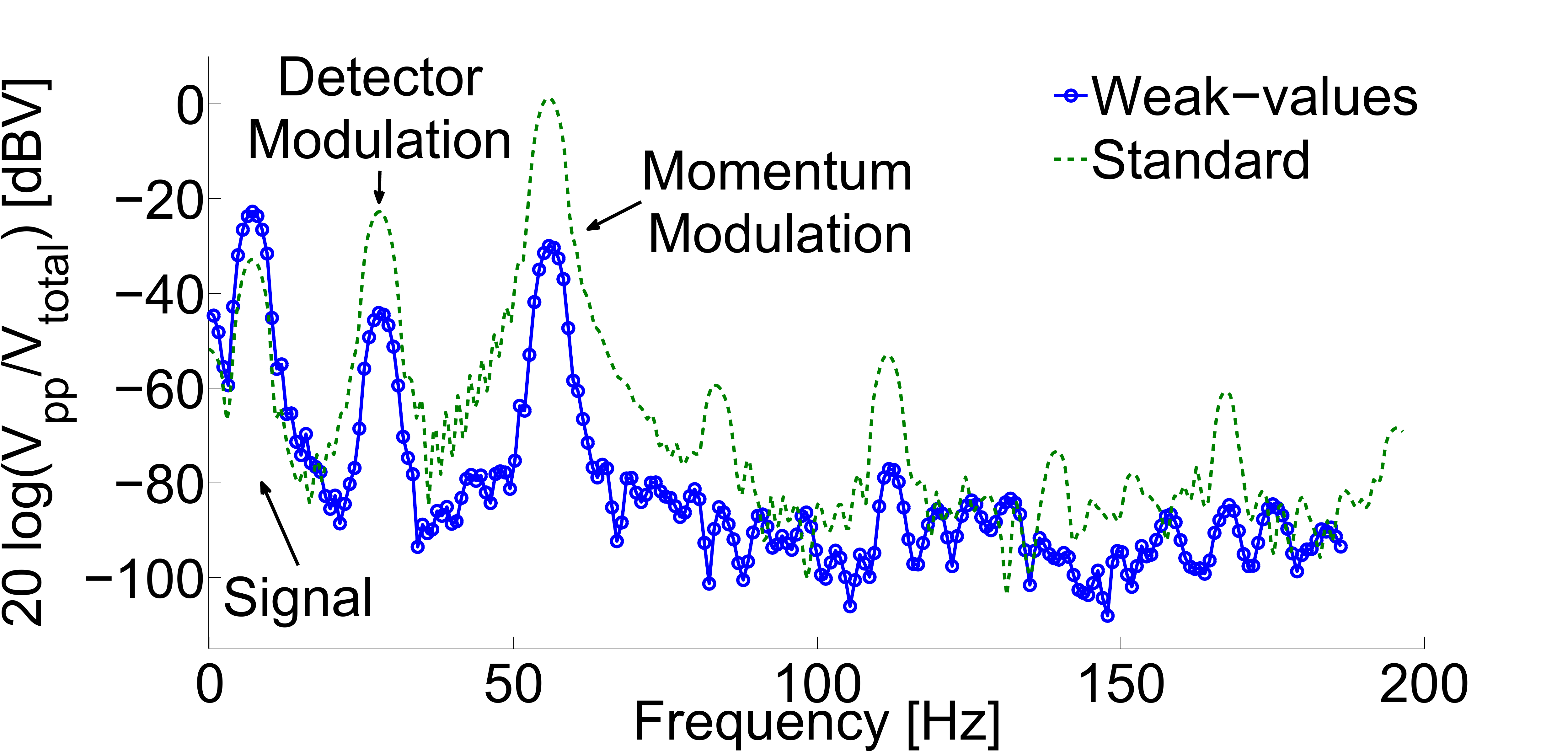}
  \caption{(Color online) A dBV spectrum comparison of the WVT (blue) and ST (green) with both external modulations, where dBV$=20\log_{10}({V/V_{total}})$ is plotted as a function of frequency. The factor $20$ in the dBV definition comes from the convention of using voltages instead of power. The peak-to-peak signals are normalized to the detected power, $V_{total}$, in their respective experiment; either the WVT or the ST. From the plot, we see a signal, $V_{pp}$, from the ``Signal" deflection corresponds to an angle of $48$ nrad peak to peak at $7$ Hz, an external transverse ``Momentum Modulation" corresponds to an angle of $2.5$ $\mu$rad peak to peak at $56$ Hz and an external transverse ``Detector Modulation" corresponds to displacement of $230$ nm peak to peak at $28$ Hz. The Fourier spectrum illustrates both the weak-value enhancement of the signal of a factor of $3.2$ over the ST, as well as the suppression of the transverse detector modulation and the transverse momentum modulation. In addition, the suppression of the two external modulations is larger than the amplification factor as predicted theoretically in Eqs.~(\ref{Eq:SpecVoltComp}). We note, the suppression of the external modulations from the signal, $V_{pp}$, collected from SD$1$ are not direct deflection measurements [see Eqs.~(\ref{Eq:SpecVoltComp}]).}
  \label{fig:AllExtMod}
\end{figure}

We now discuss measurements of $k$ in the presence of external modulations. For the WVT, we have a post-selection angle of $\phi\approx0.38$ rad and $L\approx34$ cm. The beam size is a constant $\sigma=1.075$ mm out of the fiber. For the WVT, the input power is $P^{wv}_{in}\approx1.45$ mW. In the ST, we use a focusing lens of $f=1$ m and an input power of $P^{st}_{in}\approx400$ $\mu$W. The power is lower for the ST to avoid saturating the detector. The reduction of power is accounted for by comparing the deviation to the respective lower bound so the resulting ratio is independent of the total power. Because of this, we see the WVT allows the use of more input power without saturating the detector and avoids a nonlinear response from the detector.

In Fig.~\ref{fig:AllExtMod}, the average Fourier transform of the signal measured by the SD$1$ is shown. We normalize the WVT and the ST Fourier transforms by dividing by $V_{total}$, the total voltage corresponding to the power of all detected photons in a technique. The figure can be interpreted as the visibility of the signal from each respective technique. 
The voltage $V_{pp}$ is the raw signal of the detector read by the oscilloscope. The signal from the SD$1$ on the oscilloscope is given by $V_{pp}/V_{total}=\delta x/ 2\sigma\alpha_{cal}$, where $\alpha_{cal}$ is a calibration constant of the detector and $\delta x$ is the beam displacement. We note here that the units of the Fourier transform are such that $20$ dBV is a factor of $10$ in volts. The dBV$=20\log_{10}{V/V_{total}}$. 
The signal of interest is the beam deflection labeled as ``Signal" corresponding to an angle of $48$ nrad peak to peak at $7$ Hz. The transverse ``Momentum Modulation" corresponds to an angle of $2.5$ $\mu$rad peak to peak at $56$ Hz. The transverse ``Detector Modulation" corresponds to a displacement of $230$ nm peak to peak at $28$ Hz. The transverse ``Momentum Modulation" is a piezoactuated mirror before the momentum signal $k$. The ``Detector Modulation" is the SD$1$ on a piezodriven stage. The green line is the ST with the higher harmonics of the external sources. The blue line shows the WVT with signal {\it higher} than in the ST because of the weak-value amplification. 

\begin{figure}[t!]
 \centering
 \includegraphics[scale=0.16]{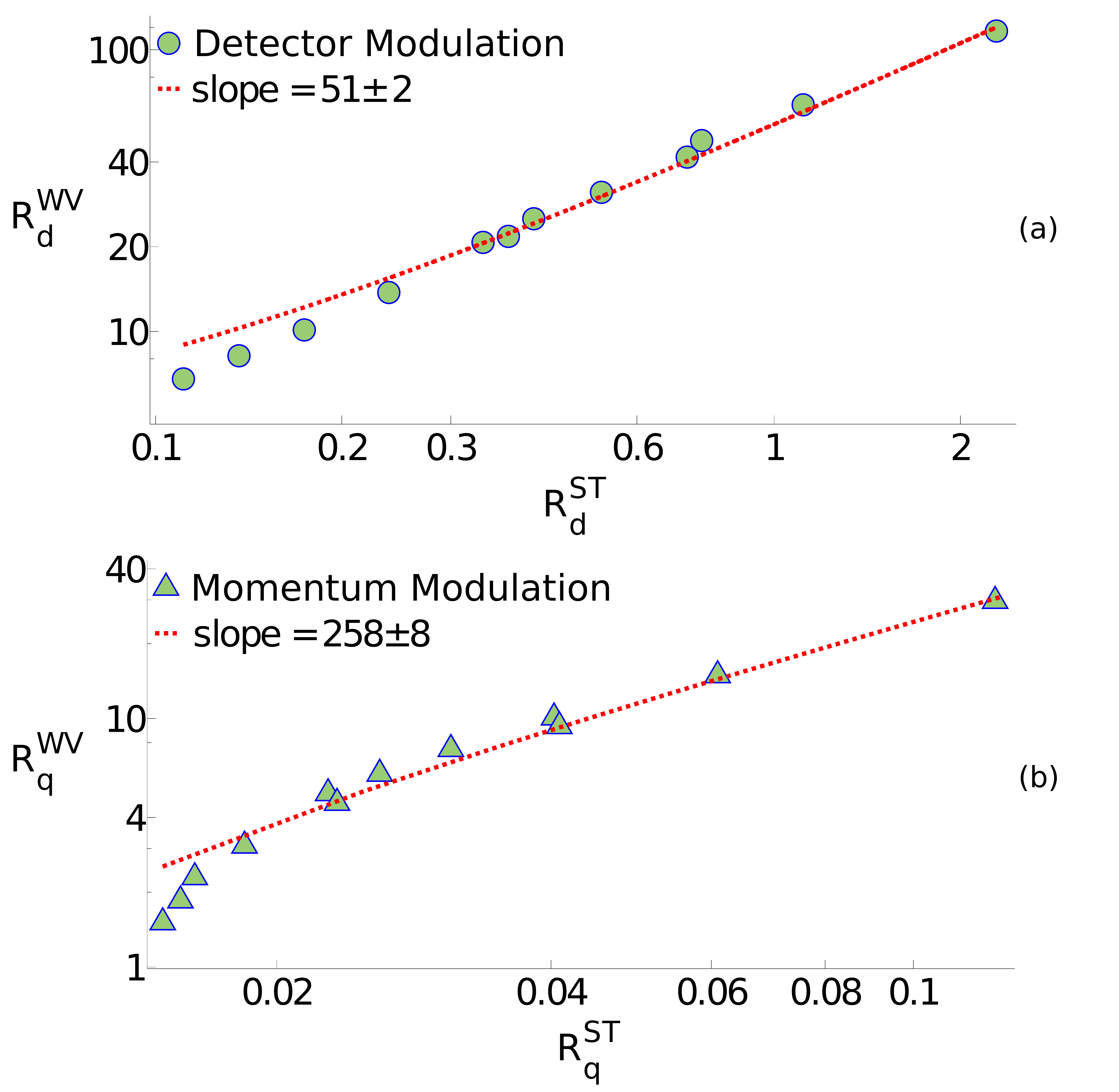}
 \caption{(Color online) A log-log plot of the ratio of the signal voltage to external modulation of the WVT, $\mathcal{R}^{wv}$, as a function of the ratio of the ST, $\mathcal{R}^{st}$, with varying external modulation strengths. In plot (a), external transverse detector modulation $d$ is applied. In plot (b), external transverse momentum modulation $q$ is applied. We use $12$ points to demonstrate the constant-slope behavior of Eqs.~(\ref{Eq:Geo}). The postselected angle is $0.38$ rad and $L\approx34$ cm. The dotted red lines are the linear fits of the data.}
 \label{fig:Geo}
\end{figure}

The spectrum analysis in Fig.~\ref{fig:AllExtMod} shows that the WVT {\it mitigates} the external modulation signals at the detector; the transverse detector modulation in volts is mitigated by $11$ times ($21$ dBV from Fig.~\ref{fig:AllExtMod}), and the transverse momentum modulation in volts is mitigated $28$ times over ($29$ dBV from Fig.~\ref{fig:AllExtMod}) the ST. We also observe a suppression of the modulations at harmonics of the driving frequencies found in the ST. The ``Signal," however, is amplified by a factor of $3.2$ ($10$ dBV from Fig.~\ref{fig:AllExtMod}) in the WVT over the ST.

The signal benefits of WVT over ST from Fig.~\ref{fig:AllExtMod} is predicted in the following (see Tab.~\ref{tab:table1}):
\begin{subequations}
\begin{equation}
\frac{\delta x^{wv}_k\sigma_f}{\delta x^{st}_k\sigma}=\cot(\phi/2),
\label{Eq:SpectrumVoltComp}
\end{equation}
\begin{equation}
\frac{\delta x^{wv}_q\sigma_f}{\delta x^{st}_q\sigma}=\frac{L}{f}\frac{\sigma_f}{\sigma}=\frac{L}{2k_0\sigma^2},
\label{Eq:SpectrumVoltComp_q}
\end{equation}
\begin{equation}
\frac{\delta x^{wv}_d\sigma_f}{\delta x^{st}_d\sigma}=\frac{\sigma_f}{\sigma}=\frac{f}{2k_0\sigma^2}.
\label{Eq:SpectrumVoltComp_d}
\end{equation}
\label{Eq:SpecVoltComp}
\end{subequations}

\noindent
We note, in the ratios in Eqs.~(\ref{Eq:SpecVoltComp}) we divide out the beam radius; so the amplification or improvement is not in accuracy, but strictly in the raw signal given by the detector.
The theoretical prediction of Eq.~(\ref{Eq:SpectrumVoltComp}) predicts the WVT amplification of $5$ over the ST ($14$ dBV for the signal $k$) for $\phi=0.38$. Likewise, the external modulation of $d$ and $q$ in the ST are $24$ and $34$ dBV, respectively, greater than the WVT.

In Fig.~\ref{fig:Geo}, we plot $\mathcal{R}$ of the WVT vs. the ST. The data are using two different $k$ values that give $48$ and $16$ nrad peak-to-peak deflections of frequency $7$ Hz. We set both external modulation sources to $28$ Hz one at a time to study them independently. By fitting the data, we arrive with the geometric factors in Eqs.~(\ref{Eq:Geo}). From these results, the WVT outperforms the ST by a factor of $258$ for transverse momentum modulations and by a factor of $51$ for transverse detector modulations for our parameters. Note the constant slope, as predicted by the theory in Eqs.~(\ref{Eq:Geo}). However, there is a discrepancy between the predicted geometric slope values of $285$ and $100$ for transverse momentum modulation and transverse detector modulation respectively. This discrepancy is consistent with previous experiments~\cite{Dixon2009,StarlingOpti} and attributed to the quality of the dark port and imperfections of the optical elements.

After verifying the theoretical behavior, we study how the deviation of $k$, $\Delta k$, is affected by the external modulations $q$ or $d$. We use a trapezoid function at frequency $10$ Hz with a rise time of $10$ ms to drive the piezoactuated BS. The trapezoid function gives a constant momentum kick for about $40$ ms. The external-modulation is a sine wave with frequency $250$ Hz and our collection window is $4$ ms. We collect data with a sample time of $T=8$ $\mu$s. This measurement protocol gives us $500$ raw data points of the momentum kick.

\begin{figure}[t!]
 \centering
 \includegraphics[scale=0.16]{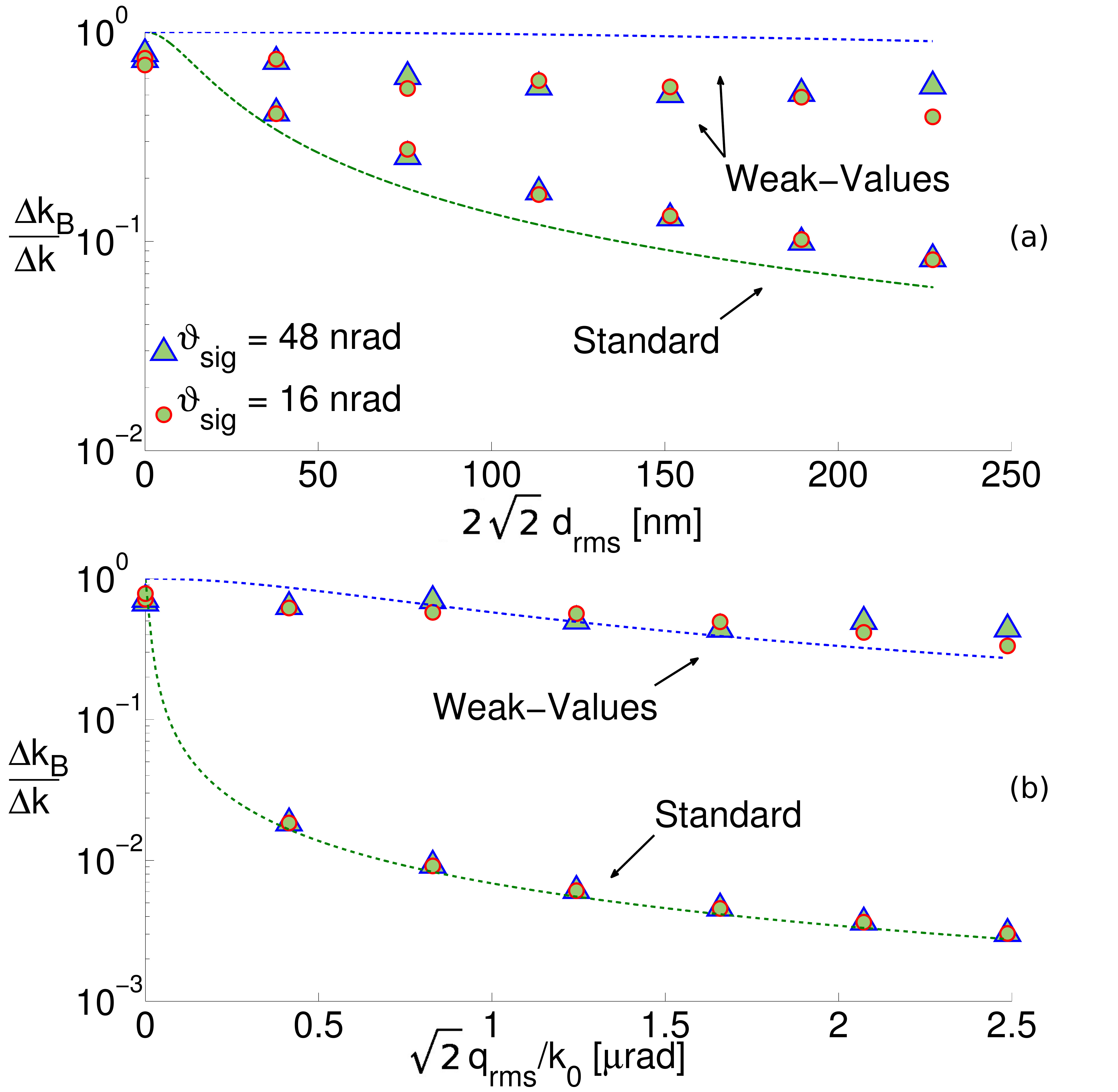}
 \caption{ (Color online) A plot of the theoretical minimum deviation given by the Cram\'{e}r-Rao bound,$\Delta k_{B}$, divided by the deviation of the measurements of $k$, $\Delta k_{B}/\Delta k=1/\sqrt{1+\xi^2_{rms}/\Delta k^2_{B}}$ as a function of external modulation strength $\xi_{rms} \in \{d_{rms},q_{rms}\}$. Data comes from a signal $k$ of $16$ and $48$ nrad deflection with variable external modulation at a frequency of $28$ Hz. Plot (a) is for transverse detector modulation and plot (b) is for transverse momentum modulation. The blue lines are the WVT theory and the green lines are the ST theory. We stress that $\xi_{rms}$ is not a noise source, but models one frequency component of a general noise source.}
 \label{fig:ExternalMod}
\end{figure}

We note the split detectors have variable gain settings with a white-Gaussian-power-dependent electronic noise, $\mathcal{J}$, equally present in both techniques.
\begin{subequations}
\begin{equation}
\mathcal{J}^{wv} = \frac{\sigma_J}{\sqrt{T}}\frac{\alpha_{cal}2\sigma}{V^{wv}_{total}}\frac{\tan(\phi/2)}{2\sigma^2}.
\end{equation}
\begin{equation}
\mathcal{J}^{st} = \frac{\sigma_J}{\sqrt{T}}\frac{\alpha_{cal}2\sigma_f}{V^{st}_{total}}\frac{k_0}{f}.
\end{equation}
\label{Eq:Jnoises}
\end{subequations}

\noindent
In Eqs.~(\ref{Eq:Jnoises}), $\sigma_J$ is the deviation of the intrinsic electrical noise (with laser off), and $T$ is the sample time. The factor $\alpha_{cal}2\sigma/V_{total}$ converts the electrical detector noise to a displacement in meters. The beam radius at the detector is defined to be $2\sigma$; $V_{total}$ is the voltage proportional to the total power on the detector, and $\alpha_{cal}\approx0.66$ is a calibration constant from the SD. The last term converts the noise to momentum units given the technique in use.

The Cram\'{e}r-Rao bound for estimating $k$ is given by $\mathcal{I}^{-1}_0$ in the absence of technical noise. So, we modify the Cram\'{e}r-Rao bound to include the uncorrelated $\mathcal{J}$ noise~\cite{Jordan2013,Knee2013} by

\begin{equation}
\Delta k^2_{B} = 1/\mathcal{I}_0+\mathcal{J}^2.
\label{Eq:kCRB}
\end{equation}
\noindent
For a fair comparison, each technique is compared to its respective lower bound in uncertainty defined by the Cram\'{e}r-Rao bound in Eq.~(\ref{Eq:kCRB}). In Fig.~\ref{fig:ExternalMod}, we plot $\Delta k_{B}$, divided by the deviation of measurements of $k$, $\Delta k=\sqrt{\Delta k^2_B + \xi^2_{rms}}$, as a function of the external modulation strength $\xi_{rms}\in\{d_{rms},q_{rms}\}$, where $\xi_{rms}$ is the root-mean-square value of the sinusoidal external modulation. When both techniques have no external modulation ($\Delta k=\Delta k_B$), $\Delta k$ is at best a factor of $7$ away from the $\mathcal{I}^{-1/2}_0$ or the shot-noise limit. 
All of the post-selection was done with $\phi\approx0.38$ rad. Fig.~\ref{fig:ExternalMod} shows the WVT is insensitive to external modulations ($1\ge\Delta k_{B}/\Delta k\ge 0.5$), while the ST is sensitive. From Figure~\ref{fig:ExternalMod}, the WVT  outperforms the ST in deviation up to a factor of $7$ for large transverse detector modulation ($230$ nm $=$ $2\sqrt{2}d_{rms}$) and $145$ for large transverse momentum modulation ($2.5$ $\mu$m $=$ $\sqrt{2}q_{rms}/k_0$).

We note, we acquire data when the signal $k$ has shifted the beam by $\delta_d$ to a steady value that remains constant for the integration time. Extracting the deviation of $k$ includes the electrical detector noise $\mathcal{J}$ and external modulation $\xi_{rms}$. We add $\Delta k^2_B$ and $\xi^2_{rms}$ in quadrature to describe the deviation of the measurement of $k$ because both sources are uncorrelated with each other. This is not a general result since one can devise a single tone modulation that will be correlated with the detector power-dependent noise and as a consequence would not be able to add the modulation in quadrature.
However, we want to stress that Fig.~\ref{fig:ExternalMod} is for a single toned external modulation uncorrelated to the power-dependent noise from the split detector. Also if one were to superimpose many of these external modulations with random frequency, phase, and amplitude one would expect behavior following the description in~\cite{Jordan2013} and not as in Fig.~\ref{fig:ExternalMod}.

Now, we discuss the effect of a Gaussian-distributed angular jitter to the Fisher information of the WVT and the ST as outlined in Ref.~\cite{Jordan2013}. The final probability distribution in position is Gaussian distributed and the ST has a mean $kf/k_0$ and variance $\sigma^2_f+f^2Q^2/k^2_0$, where $Q^2$ is the angular-jitter variance. The probability distribution for the WVT has mean $2k\sigma^2\cot(\phi/2)$ and variance $\sigma^2+(L/2k_0\sigma)^2(1+(2\sigma Q)^2)$. The Fisher information for both techniques is given by:

\begin{subequations}
\begin{equation}
\mathcal{I}_{Q}^{wv}(k) = \frac{4N\sigma^2}{1+(\frac{L}{2k_0\sigma^2})^2[1+(2\sigma Q)^2]},
\label{Eq:InfoModWV}
\end{equation}
\begin{equation}
\mathcal{I}_Q^{st}(k) = \frac{4N\sigma^2}{1+(2\sigma Q)^2},
\label{Eq:InfoModST}
\end{equation}
\label{Eq:InfoMod}
\end{subequations}

\noindent
where the subscript $Q$ denotes the angular-jitter analysis.

The Fisher information for the WVT in Eq.~(\ref{Eq:InfoModWV}) shows suppression of the angular jitter with larger $\sigma$ and with shorter $L$, the distance from the source $Q$ to detector. However, the Fisher information for the ST in Eq.~(\ref{Eq:InfoModST}) degrades as $\sigma_f$ decreases ($\sigma_f\propto 1/\sigma$). From the Fisher information perspective, it is better to use a long focus to acquire more of the available Fisher information in the ST, but this will introduce turbulence effects~\cite{Jordan2013}.

We also discuss the effect of a Gaussian-distributed detector-displacement jitter to the Fisher information of the WVT and the ST as outlined in Ref.~\cite{Jordan2013}. 
If the detector-displacement jitter has a variance of $J^2$ then the ST variance at the detector becomes $\sigma_f^2+J^2$ and the WVT variance becomes $\sigma^2+J^2$ such that the Fisher information for both techniques is given by:
\begin{subequations}
\begin{equation}
\mathcal{I}_{J}^{wv}(k) = \frac{4N\sigma^4}{\sigma^2+J^2},
\label{Eq:InfoModWVdetect}
\end{equation}
\begin{equation}
\mathcal{I}_{J}^{st}(k) = \frac{N(f/k_0)^2}{\sigma_f^2+J^2}=\frac{4N\sigma^2}{1+(\frac{2k_0\sigma J}{f})^2}.
\label{Eq:InfoModSTdetect}
\end{equation}
\label{Eq:InfoModdetect}
\end{subequations}

\noindent
Where the subscript $J$ denotes the detector-displacement jitter analysis. This symbol is not to be confused with $\mathcal{J}$, the electrical noise on the detector, from Eqs.~(\ref{Eq:Jnoises}).

Similarly, the Fisher information in Eq.~(\ref{Eq:InfoModWVdetect}) shows suppression of the detector jitter with large $\sigma$ such that $\sigma\gg J$, while the ST Fisher information in Eq.~(\ref{Eq:InfoModSTdetect}) shows to be optimal with detector-displacement jitter only for large values of focal length such that $f\gg 2k_0\sigma J$. This is the same as having a larger displacement at the detector ($\sigma_f\gg J$) but will introduce turbulence effects~\cite{Jordan2013}. Thus, the WVT with detector-displacement jitter will outperform the ST under the Fisher information metric.

The Cram\'{e}r-Rao bound derived from the Fisher information of both angular jitter Eqs.~(\ref{Eq:InfoMod}) and beam-displacement jitter Eqs.~(\ref{Eq:InfoModdetect}) leads to a similar behavior to our external modulation results in Figs.~\ref{fig:ExternalMod}b and~\ref{fig:ExternalMod}a, respectively. This analysis for the Gaussian-distributed noises from~\cite{Jordan2013} reveals that the WVT is superior over the ST in obtaining Fisher information with technical noise. Our experimental results only encompass one frequency component in the theory but validate the behavior.

\begin{figure}[t!]
  \centering
  \includegraphics[scale=0.15]{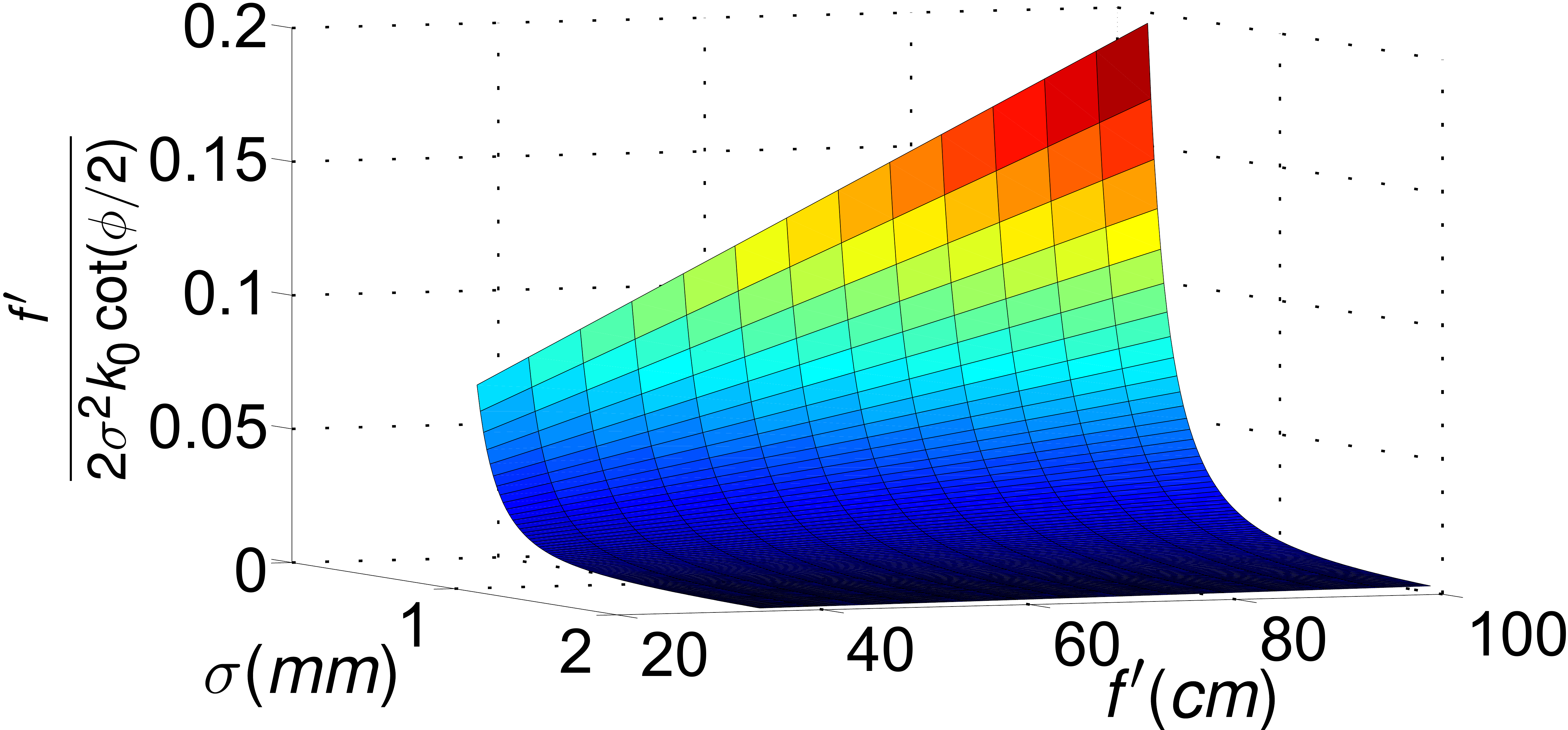}
  \caption{(Color online) A plot of the geometric factor $f'/2\sigma^2k_0\cot(\phi/2)$ from Eq.~(\ref{Eq:Geo}) as a function of beam radius $\sigma$ and focal length $f'=f$ or distance $f'=L$. The plot never surpasses $1$, thus the ST will not outperform the WVT. The plot uses a postselection angle of $\phi=0.4$ rad.}
  \label{fig:Geo3d}
\end{figure}

When comparing each technique with external modulations as in Eqs.~(\ref{Eq:Geo}) the WVT \emph{always} outperforms the ST. We explore possible parameter space to \emph{reoptimize} the ST with the following assumptions. (i) Assuming both $f$ and $L$ to be no greater than one meter to avoid turbulence (as discussed in Ref.~\cite{Jordan2013}).  (ii) We fix the maximum value of $\sigma$ no more than $2$ $mm$ and no smaller than $250$ $\mu m$ (to avoid saturation of the detector).

In Fig.~\ref{fig:Geo3d}, we plot the geometric factor of Eqs.~(\ref{Eq:Geo}) as $f'/2\sigma^2k_0\cot(\phi/2)$ for experimentally possible parameters of $f'$ and $\sigma$, where $f'$ can be either $f$ or $L$. The postselection angle for the plot is $\phi=0.4$ rad. From the figure, the geometric value never exceeds $1$. Thus in this comparison the WVT \emph{always} outperforms the ST. From Eqs.~(\ref{Eq:Geo}) and Fig.~\ref{fig:Geo3d}, the WVT advantage over external modulations increases for smaller postselection angles and larger beam widths~\footnote{If we consider the bright port case the $\cot(\phi/2)$ in Eqs.~(\ref{Eq:Geo}) will change to $\tan(\phi/2)$. From this we conclude that when considering technical noise only the dark port of the WVT outperforms the ST.}. Thus, the technical advantage of the WVT is controlled by the geometry and parameter selection for the experiment.

\section{Results---Efficiency of Estimation}\label{Results2}
Next, we study the efficiency of the estimator by using the Fisher information in absence of external modulations. To extract the Fisher information behavior predicted in Eqs.~(\ref{Eq:FisherWV}), we collected the photons from both bright and dark ports. As pointed out in Refs.~\cite{Jordan2013,Knee2013,Ferrie2013}, the bright port in general also has information about the parameter in it. 
Instead of using a trapezoid wave, here we used a $7$ Hz sine wave for the momentum kick $k$ and varied the postselection angle. Then, we measured momentum kick $k$ and the deviation, $\Delta k$, from both the dark and bright ports with SD$1$ and SD$2$, respectively (see Fig.~\ref{fig:setup}). Averaging the Fourier transform of the signal allowed us to extract the SNR, $\mathcal{S}$.  We acquired data from the Fourier transform of the signal and note the procedure is only affected by the component of $\mathcal{J}$ of the same frequency as the signal $k$. For this uncorrelated temporal Gaussian noise, the Fisher information is related to the SNR as:
\begin{figure}[b!]
 \centering
 \includegraphics[scale=0.16]{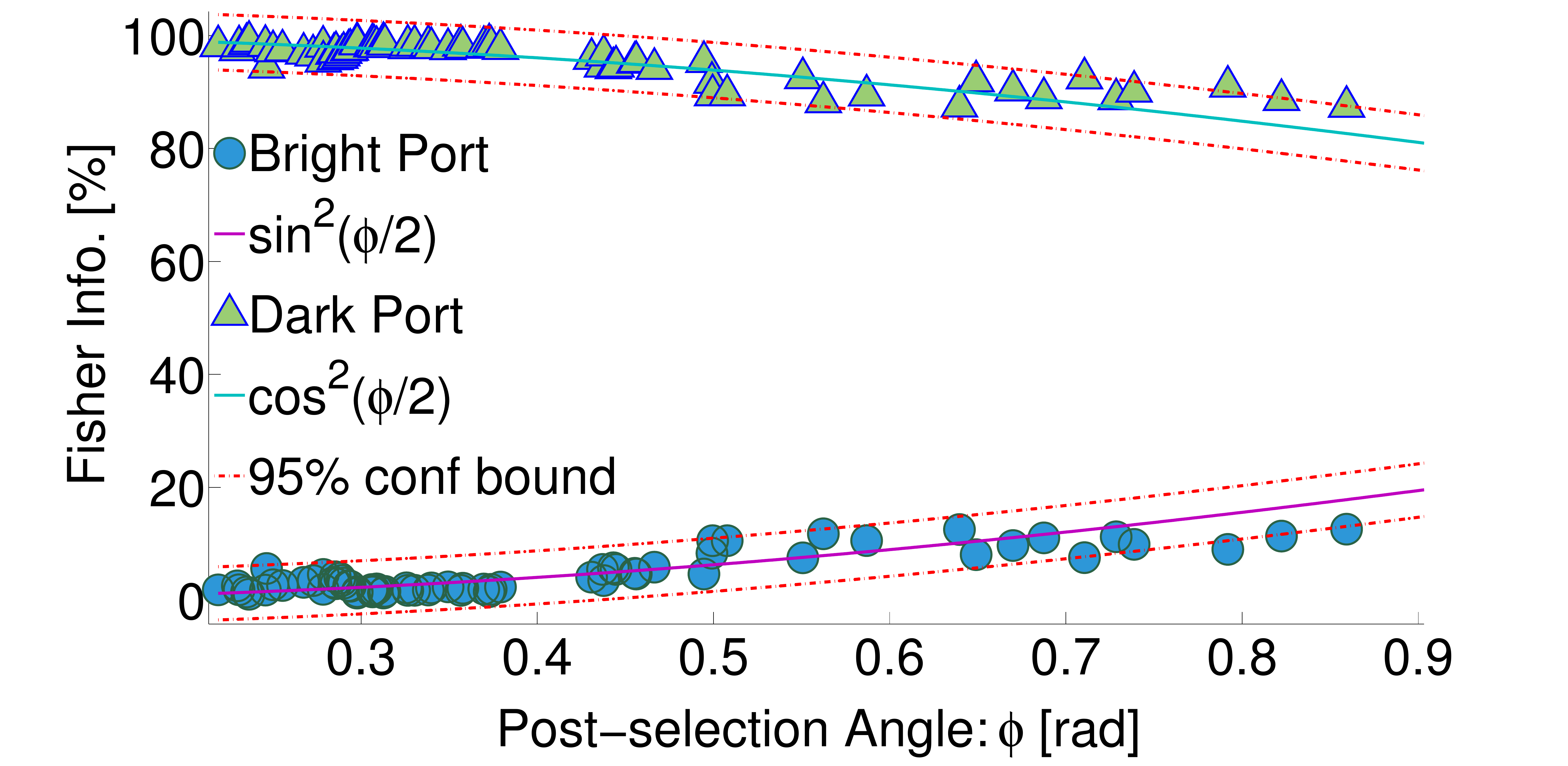}
 \caption{(Color online) Fisher information vs post-selection angle $\phi$. Data is taken from the Fourier transform and averaged over equal numbers of samples. Angle $\phi$ ranges from $0.22$ to $0.9$ rad. The confidence interval is $95\%$, and we see the fit break down as $\phi$ becomes large. Most of the information is found to be in the dark port even for large $\phi$. Both dark and bright ports follow $\cos^2(\phi/2)$ and $\sin^2(\phi/2)$ behavior, respectively, as in Eqs.~(\ref{Eq:FisherWV}).}
 \label{fig:FisherInfo}
\end{figure}
\begin{equation}
\mathcal{S}^2=\left(\frac{k}{\Delta k_{B}}\right)^2=k^2\mathcal{I}.
\end{equation}
Since both bright and dark ports are measuring the same $k$, we arrive at the percentage of Fisher information from each port given the total Fisher information available,
\begin{equation}
\mathcal{I}^{\%}_{D,B} = \frac{\mathcal{S}^2_{D,B}}{\mathcal{S}^2_D+\mathcal{S}^2_B}.
\label{Eq:FisherPercent}
\end{equation}
Here, we define $\mathcal{I}^{\%}_{D,B}$ as the percentage of Fisher information in dark ($D$) or bright ($B$) ports.

In Fig.~\ref{fig:FisherInfo}, the percentages of Fisher information from each port are shown as a function of postselection angle. We observe the corresponding behavior of the weak-value regime of Eqs.~(\ref{Eq:FisherWV}) and note that most of the information is recovered from the dark port with a small post-selection angle $\phi$. We fit the data with about $100$ points for both dark and bright ports. The Fisher information is a near-perfect match to the theoretical prediction. The nonlinear fit gives a goodness measure $r^2=0.99$, and the red lines are the $95\%$ confidence interval bounds ($2\sigma_{error}$). Note that the results deviate from the approximation as $\phi$ increases out of the weak interaction approximation of Eq.~(\ref{Eq:wvOut}). 
In addition, we find $99\pm2\%$ of the Fisher information in the dark port and $1\pm2\%$ of the Fisher information in the bright port for a postselection angle of $\phi\approx0.22$  rad ($1\%$ of the photons). Even though we only measure $1\%$ of the photons, we extract $99\%$ of the Fisher information. From the results we conclude weak-value amplification with strong postselection (dark port) extracts almost $100\%$ of the Fisher information about the momentum kick $k$, while the Fisher information in weak value amplification with failed post-selection (bright port) is negligible for practical purposes. As predicted in Eqs.~(\ref{Eq:FisherWV}) and~(\ref{Eq:StFisher})~\cite{Jordan2013}, the weak-value amplification technique provides an efficient estimation for this experiment. We note that using an estimator that also incorporates the bright port will make this technique {\it even better}, but only slightly.

Although we have extracted $99\%$ of the Fisher information from $1\%$ of the photons, we wish to stress that this is in no way a limit on the efficiency of the technique, but a proof-of-principle result.  We can quantify this point in the following manner:  Suppose we wish to demonstrate the efficiency of the weak-value estimator explored in this paper to some fixed fraction of the total Fisher information, $1-\epsilon$, where $\epsilon$ is a small, but finite number.
This is equivalent to showing ${\mathcal{I}_D^\%} > 1-\epsilon$.
We can demonstrate the efficiency of the technique to this level by fixing the post-selection angle to be
\begin{equation}
\phi/2 <  \sqrt{\epsilon},
\label{Eq:result}
\end{equation}
where we recall the fractional Fisher information Eqs.~(\ref{Eq:FisherPercent}) and~(\ref{Eq:FisherWV}) in this experiment~\cite{Pang,Jordan2013}.
This assumes $k^2\sigma^2\cot(\phi/2)\ll1$ (controlling the weakness of the interaction) is suitably reduced as well, while also measuring a sufficiently large number of photons. Amazingly, Eq.~(\ref{Eq:result}) indicates that the technique is {\it more} efficient the {\it fewer} photons measured in the dark port. Since $\epsilon$ can be made small, we conclude the technique can be made as efficient as desired in principle. The important practical limitation is the fidelity of the optics, getting a good dark port, and any other deviations from the theory.

In the paper by Jordan \emph{et al}.~\cite{Jordan2013} the theoretical predictions are in the classical Fisher information but the result is the same as quantum Fisher information. Both techniques are near shot-noise limited, (see Fig.~\ref{fig:ExternalMod}), and the techniques are assumed to reach their respective Cram\'{e}r-Rao bound~\cite{Viza,Steinberg,Nishizawa,Jordan2013}. Therefore, our results in Fig.~\ref{fig:FisherInfo} gives strong evidence that the WVT will perform the same if we send individual photons through the experiment. From a metrological perspective the precision of $N$ independent measurements is recovered from only a fraction of measurements. Even from the classical perspective as we throw away $99\%$ of the intensity in the WVT we recover the precision associated with using all the available measurements as in the ST. Thus the WVT in a sense squeezes \emph{all} the available Fisher information to the few surviving postselected photons.



\section{Results---Noise in the Wild}\label{Results3}
\begin{figure}[t!]
 \centering
 \includegraphics[scale=0.16]{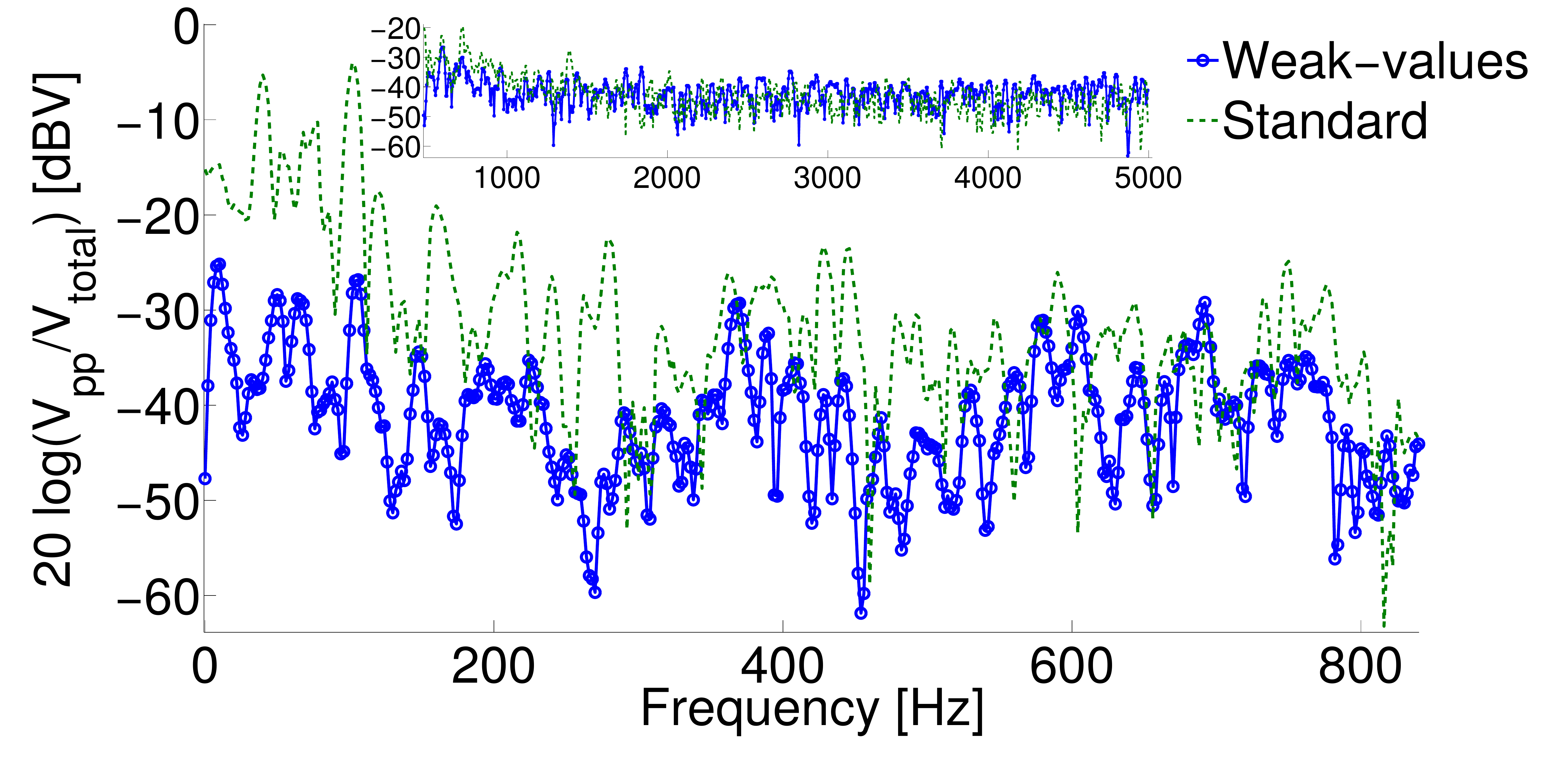}
 \caption{(Color online) A spectrum voltage comparison of the WVT (blue) and ST (green) with naturally occurring laser-beam-jitter noise, where $V_{pp}$ is the signal from the detector in volts. Without the fiber, the beam is shaped to $\sigma = 1.12$ mm and is sent to the experiments. The beam jitter is found in the low frequency regime (under $1$ kHz). We see about $20$ dBV improvement in the WVT for low frequencies (under $300$ Hz). We note similarly to Fig.~\ref{fig:AllExtMod}, the comparison is not of beam shift, but of voltages from the laser-beam-jitter noise. Both data sets were taken by averaging 128 samples. The plots are normalized to the detected power, $V_{total}$ (either the WVT or the ST in their respective experiments). Note that this is a voltage comparison and not a deflection comparison of WVT and the ST.}
 \label{fig:BeamJitterFreq}
\end{figure}

In Fig.~\ref{fig:AllExtMod}, the amplitude of the angular modulation outside the interferometer is suppressed in the WVT, relative to the ST.  This behavior was predicted theoretically to occur, regardless of the frequency of the oscillation \cite{Jordan2013}. We will now see how this effect can be put to use for a more general noise in the wild. To accomplish this, we removed the connecting fiber that stabilizes the laser, and direct the light into one of the two experiments in Fig.~\ref{fig:setup}.  The signal on the detector then registers noise that is a combination of electronic noise and intrinsic laser jitter. We note that the statistics of this jitter is neither white, nor Gaussian, nor is it stationary. The angular jitter originates from the physics of the laser, and exists up to around $300$ Hz in this experiment. It has strong frequency components at around $50$ and $100$ Hz.  Its constantly changing statistical nature makes any kind of improved statistical estimation strategy extremely challenging. Nevertheless, the fact that the weak-value experiment globally suppresses the amplitude of all angular jitter from outside the interferometer makes the WVT very convenient as a noise reduction strategy. Indeed, we see from Fig.~\ref{fig:BeamJitterFreq} that the contribution of the laser jitter to the noise spectrum is essentially eliminated entirely, being reduced below the electronic noise floor.

In Fig.~\ref{fig:BeamJitterFreq}, the Fourier transform of both the WVT (blue) and the ST (green) signal as a function of frequency is given. The Fourier transforms shown are the average of $128$ samples and the WVT postselection angle is $\phi\approx0.46$ rad. We note, while the ST uses $400$ $\mu$W and the WVT uses $1.45$ mW of power, the Fourier transform of the signals of both are renormalized given the total detected power used in each technique for a fair comparison.

Next, we made the measurements in the time domain with a sample time of $T=4$ ms and compared the relative error of $k$ in both techniques. The relative error is the deviation of the measurements of $k$, $\Delta k$, divided by its respective lower bound, $\Delta k_{B}$ from Eq.~(\ref{Eq:kCRB}). The relative error of the ST is $144$ and the WVT is $5$. Therefore, WVT suppresses intrinsic beam-jitter noise at best $29$ times over the ST. Most importantly from Fig.~\ref{fig:BeamJitterFreq}, the WVT completely suppresses this laser-beam-jitter noise, showing only electronic noise from the detector.

We independently verified the intrinsic laser beam jitter to be about $0.3$ $\mu$rad peak to peak using the full width at half maximum and twice the deviation of the data collected from Fig.~\ref{fig:BeamJitterFreq}. The WVT has a total propagation length of $205$ cm from the laser to detector. The ST used a focal length of $1$ m. We can verify the claim that the WVT globally suppresses laser-beam-jitter noise by comparing the suppression of the intrinsic (stochastic) beam jitter to the single frequency modulation at an amplitude chosen to be the typical wander. From the data in Fig.~\ref{fig:ExternalMod}(b) where one single frequency is modulating an external mirror before the interferometer (see Fig.~\ref{fig:setup}), we can predict what the mitigation factor is for a single tone of deflection angle $0.3$ $\mu$rad. According to Fig.~\ref{fig:BeamJitterFreq}, the suppression factor for the intrinsic beam jitter is at best $29$ and the suppression for the single-frequency tone from Fig.~\ref{fig:ExternalMod}b is $44$ (at $0.3$ $\mu$rad peak to peak), giving comparable results.


\section{Conclusions}\label{Conclusions}
This paper has been focused on two major issues. The first is a comparison of two experimental techniques, ST and WVT. The ST is a standard angle deflection technique off of a tilted mirror, while the WVT includes a beam splitter, making the system an interferometer that may be interpreted as a realization of the Aharonov, Albert, Vaidman, weak-value amplification effect if one output port is monitored \cite{Aharonov}. In the absence of any technical limitations, it is important to stress that both systems give the same fundamental limitation [see Eqs.~(\ref{Eq:FisherWV}) and Eq.~(\ref{Eq:StFisher})] on the measurement uncertainly of the mirror tilt, given the same number of input photons. Therefore, the ``weak-value amplification" alone gives no metrological advantage, unless it is combined with the other effects we have identified. 
We have noted this point some time ago \cite{StarlingOpti}, though some authors have recently rediscovered it \cite{Ferrie2013,Tanaka,Zhu} and included the study of pixelation and uncorrelated transverse jitter~\cite{Knee2013}.  However, under realistic conditions such as a detector that saturates (responds nonlinearly), the presence of vibrational detector noise or the presence of angular jitter, we have shown the WVT can perform orders of magnitude better than the ST. This is consistent with independent investigations using variations of this experiment, claiming record precision~\cite{Kasevich,Turner}. We have reported experimental results quantifying this effect under the presence of transverse detector modulations and transverse momentum modulations.

The second major issue considered in this paper is an analysis of how well a given experimental technique---the weak-value-based experiment---uses the available information contained in the data.  This checks just how efficient the weak-value-based technique is, in light of criticisms that neglecting other information sources by the postselection makes the metrological technique inefficient \cite{Knee2013,Ferrie2013,Tanaka,Zhu}. We have demonstrated experimentally that by measuring only $1\%$ of the light in the experiment, $99\%$ of the theoretically available information may be extracted from it, as we have theoretically predicted \cite{Jordan2013}. In principle, the remaining $1\%$ of the information can be extracted from the bright port. However, the corresponding signal deamplification makes the problem one of finding a small signal in a bright background technically difficult. We have also shown how the efficiency can be further boosted by measuring a {\it smaller} fraction of the photons if desired, and consequently a well-designed weak-value-based metrology experiment is remarkably efficient. In a sense, the WVT can be viewed as a filtering procedure where the selected photons carry the vast majority of the Fisher information, and the noninformative photons have been filtered out.

When combined with other ideas of signal recycling~\cite{Kasevich} and power recycling \cite{DresselRecycle,Kevin}, or quantum enhancements~\cite{Pang,Dressel2}, we anticipate future weak-value-based metrological experiments will be able to reach even greater levels of precision.

\begin{acknowledgements}
We thank J. Dressel, C. Ferrie, and G. C. Knee, for helpful comments on the manuscript. We give additional thanks for the careful editing from J. Schneeloch and B. J. Little. This work was supported by the Army Research Office Grant No. W911NF-12-1-0263 and No. W911NF- 09-0-01417, as well as the CAPES Foundation, process No. BEX 8257/13-2.
\end{acknowledgements}

%



\end{document}